# Parametric Modeling of EEG by Mono-Component Non-Stationary Signal

Pradip Sircar* and Rakesh Kumar Sharma

Department of Electrical Engineering

Indian Institute of Technology Kanpur

Kanpur 208016, Uttar Pradesh, India

*Corresponding Author, Email: sircar@iitk.ac.in

*Abstract*

**In this paper, we propose a novel approach for parametric modeling of electroencephalographic (EEG) signals. It is demonstrated that the EEG signal is a mono-component non-stationary signal whose amplitude and phase (frequency) can be expressed as functions of time. We present detailed strategy for estimation of the parameters of the proposed model with high accuracy. Simulation study illustrates the procedure of model fitting. Some interpretation of the characteristic features of the model is described.**

*Keywords-* Parametric modeling, Non-stationary model, EEG signal, Evolutionary signal, Amplitude modulated signal, Phase modulated signal

## 1. Introduction

The study of electroencephalographic (EEG) signals provides a non-invasive means for revealing the functional states of the brain and for diagnosing functional brain disturbances [1, 2]. The primary purpose of analysis of the EEG signals is to extract information from the recording, which will complement the visual evaluation of a neurophysiologist. The EEG signals representing brain activity are multi-featured signals with variety of information, and considering the inherent complexity of the signals, researchers employed diversified linear and nonlinear techniques for extraction, classification, and interpretation of EEG features related to the conditions and dysfunctions of brain [3, 4].

A document search on SCOPUS for the word "EEG" yields more than 100,000 results till the end of the year 2019, with an average rate of publication over 6–8 thousand per year in recent time, and the rate is steadily increasing. A wide variety of techniques of signal processing and soft computing have been employed for analysis, information extraction and classification of the EEG signals. Clearly, a review of the vast literature on electroencephalogram (EEG) is beyond the scope of the present paper. The mystery about the EEG signals does not seem to be resolved.

Traditionally, the EEG signal analysis by spectral decomposition reveals rhythmical repetitive waves of varied frequency and amplitude ranges [5]. The alpha and beta rhythms appear in adults during wakefulness, when mind is relaxed or active, respectively, whereas the theta and delta rhythms manifest during drowsiness and sleep, respectively [6]. Apart from the rhythmical activities of the EEG, we often observe various complex patterns, e.g., spikes or sharp waves, in the EEG recording. These observed complex patterns have significance in diagnosis of epilepsy or seizure disorders [2, 5].

Incessant variations of frequency and amplitude ranges of the EEG signals make the signals dominantly non-stationary in nature [7, 8]. Rapid changes of frequency amplitude values over short span of time are depicted in the signals as transients, e.g., spikes or sharp waves. Note that the spectral analysis of the EEG signals over some duration of time reveals all the frequencies that manifest over the duration. However, the spectral analysis does not provide any indication about the time instants when a specific band of frequency appears or disappears. Considering the non-stationary nature of the EEG signals, multi-resolution analysis of the rhythms of EEG provides better insight of brain activity [9−11].

The spectral analysis of the EEG using the parametric time-series modeling provides better accuracy than by the non-parametric approach based on the Fourier transform, specially so when the length of the data record is short [12]. However, it should be emphasized that the constant-coefficient time-series model does not provide the evolutionary nature of the EEG spectrum. Therefore, in our search for a parametric model for the EEG signal, we do not consider the time-series model to be suitable.

In this paper, we postulate that the EEG signal is a mono-component non-stationary signal, and the signal can be represented by the real part of an analytical signal of the form $A(t)\exp[j\varphi(t)]$, $A(t) \geq 0$ [13, 14]. A mono-component signal is represented as the trajectory of a single point, or the trajectory of the center of a single mountain ridge on the time-frequency plane [13]. The instantaneous angular frequency of the signal is defined by $\omega(t) = \varphi'(t)$ [14].

In general, the amplitude function $A(t)$ and the phase function $\varphi(t)$ can be arbitrary, except that $A(t)$ must be non-negative. However, without loss of generality, we will assume that the signal $A(t)\cos[\varphi(t)]$ can be represented locally as the amplitude modulated (AM) and phase modulated (PM) sinusoidal signal, with single-tone modulations in amplitude and phase terms [15, 16]. In this way, we can develop an appropriate method for estimating all the parameters of the model locally, and then, the global model is found by fitting the local parameters in the polynomials of time variable. In other words, the EEG signal will be parameterized by the simplest model locally, and the evolutionary nature of the signal will be captured globally by introducing time variations of the model parameters.

## 2. Amplitude and Phase Modulated Sinusoidal Signal Model

Let the discrete-time random signal $x[n]$ be represented as given below:

$$x[n] = A\cos[\omega_c n + k_p \sin(\omega_p n) + \theta] + \frac{sAk_a}{2}\cos[\omega_c n + \omega_a n + k_p \sin(\omega_p n) + \theta + \theta_a]$$
$$+ \frac{rAk_a}{2}\cos[\omega_c n - \omega_a n + k_p \sin(\omega_p n) + \theta - \theta_a - \theta_b] \quad (1)$$

where $A$ is the amplitude and $\theta$ is the random phase of the carrier wave, $\theta$ is assumed to be uniformly distributed over $[0, 2\pi)$, $\omega_c = 2\pi f_c$ is the carrier angular frequency,

$\omega_a = 2\pi f_a$ is the modulating angular frequency and $k_a$ is modulation index of the AM signal, $\omega_p = 2\pi f_p$ is the modulating angular frequency and $k_p$ is the modulation index of the PM signal, $\theta_a$ is the phase of the AM signal, $\theta_b$ is the additional phase of the lower side band, $s$ is the scaling factor for the upper side band, $r$ is the scaling factor for the lower side band, and $n$ is the time index.

It should be noted that the signal model $x[n]$ is the amplitude and phase modulated sinusoidal (APMS) signal. Moreover, each parameter $\xi$ of the APMS signal of (1) are assumed to be time-dependent in the form of a polynomial $\beta(n) = c_0 + c_1 n + c_2 n^2 + \cdots$, where $c_0, c_1, c_2, \cdots$ are constants. The time dependence of each parameter is not shown explicitly in (1) for clarity. Furthermore, it is assumed that when the time duration of $x[n]$ is appropriately restricted, each parameter can be assumed to be constant over the duration. Note that in this way, we can develop a time-variant global model with simple local estimation technique. Without loss of generality, the scaling factor for the lower side band is assumed to be $r = -1$ [15].

**2.1 Autocorrelation Function**

Setting the scaling factors $s = r = 1$ and the additional phase $\theta_b = 0$, (1) reduces to

$$A\left[1 + k_a \cos(\omega_a n + \theta_a)\right] \cos\left[\omega_c n + k_p \sin(\omega_p n) + \theta\right] \quad (2)$$

which is the basic APMS signal model that we will employ to represent the EEG signal. The time-variant autocorrelation function (ACF) $r_x$ of the signal in (2) is calculated as

$$\begin{aligned} r_x[n,m] &= \mathrm{E}\{x[n]x[n+m]\} \\ &= \frac{A^2}{2} \left\{ \begin{aligned} &1 + k_a \cos[\omega_a n + \theta_a] + k_a \cos[\omega_a (n+m) + \theta_a] \\ &+ \frac{k_a^2}{2} \cos[\omega_a (2n+m) + 2\theta_a] + \frac{k_a^2}{2} \cos[\omega_a m] \end{aligned} \right\} \\ &\quad \times \cos\left\{\omega_c m + k_p \sin[\omega_p (n+m)] - k_p \sin[\omega_p n]\right\} \end{aligned} \quad (3)$$

where $\mathrm{E}$ stands for the expectation operation with respect to the random variable $\theta$. The APMS signal in (2) or in (1) is a non-stationary signal, as shown by the time-variant ACF

$r_x$ in (3), which depends on both of the variables of time $n$ and lag $m$. In fact, the double dependence of ACF makes it unsuitable for use in the estimation of model parameters. Another point to be noted is that the ensemble average of (3) cannot be replaced by the time average when a single observation of the random signal $x[n]$ is available. For a single realization of $x[n]$, the phase $\theta$ is a constant.

## 2.2 Product Function

To estimate the parameters of the APMS signal in (1), we compute the product function $p_x[m]$ defined as [17]

$$p_x[m] = x[n]x[n+m]\big|_{n=-m/2}, \quad m = 0, \pm 2, \pm 4, \cdots \qquad (4)$$

Evaluating (4) with substitution of variable $l = m/2$, with $l = 0, \pm 1, \pm 2, \cdots$, we get

$$\begin{aligned}
p_x[l] =\ & \frac{A^2}{2}\cos[2\theta] + \frac{A^2}{2}\cos[2\omega_c l + 2k_p \sin(\omega_p l)] \\
& + \frac{sA^2 k_a}{4}\cos[\omega_a l - 2\theta - \theta_a] + \frac{sA^2 k_a}{4}\cos[2\omega_c l + \omega_a l + 2k_p \sin(\omega_p l) - \theta_a] \\
& + \frac{rsA^2 k_a^2}{8}\cos[2\omega_a l - 2\theta + \theta_b] + \frac{rsA^2 k_a^2}{8}\cos[2\omega_c l + 2k_p \sin(\omega_p l) - 2\theta_a - \theta_b] \\
& + \frac{rA^2 k_a}{4}\cos[\omega_a l + 2\theta - \theta_a - \theta_b] + \frac{rA^2 k_a}{4}\cos[2\omega_c l - \omega_a l + 2k_p \sin(\omega_p l) + \theta_a + \theta_b] \\
& + \frac{sA^2 k_a}{4}\cos[\omega_a l + 2\theta + \theta_a] + \frac{sA^2 k_a}{4}\cos[2\omega_c l + \omega_a l + 2k_p \sin(\omega_p l) + \theta_a] \\
& + \frac{s^2 A^2 k_a^2}{8}\cos[2\theta + 2\theta_a] + \frac{s^2 A^2 k_a^2}{8}\cos[2\omega_c l + 2\omega_a l + 2k_p \sin(\omega_p l)] \\
& + \frac{rA^2 k_a}{4}\cos[\omega_a l - 2\theta + \theta_a + \theta_b] + \frac{rA^2 k_a}{4}\cos[2\omega_c l - \omega_a l + 2k_p \sin(\omega_p l) - \theta_a - \theta_b] \\
& + \frac{r^2 A^2 k_a^2}{8}\cos[2\theta - 2\theta_a - 2\theta_b] + \frac{r^2 A^2 k_a^2}{8}\cos[2\omega_c l - 2\omega_a l + 2k_p \sin(\omega_p l)] \\
& + \frac{rsA^2 k_a^2}{8}\cos[2\omega_a l + 2\theta - \theta_b] + \frac{rsA^2 k_a^2}{8}\cos[2\omega_c l + 2k_p \sin(\omega_p l) + 2\theta_a + \theta_b]
\end{aligned} \qquad (5)$$

A close look at (5) reveals that the spectrum of sequence $p_x[l]$ will contain clusters of peaks centered at angular frequencies $2\omega_c$, $(2\omega_c - \omega_a)$, $(2\omega_c + \omega_a)$, $(2\omega_c - 2\omega_a)$,

$(2\omega_c + 2\omega_a)$, together with peaks at angular frequencies $\omega_a$ and $2\omega_a$, and a dc component. Note that by using the identity [18]

$$e^{jx\sin\varphi} = \sum_{m=-\infty}^{\infty} J_m(x) e^{jm\varphi} \tag{6}$$

or, separating the real and imaginary parts,

$$\begin{aligned}\cos(x\sin\varphi) &= \sum_{m=-\infty}^{\infty} J_m(x)\cos(m\varphi), \\ \sin(x\sin\varphi) &= \sum_{m=-\infty}^{\infty} J_m(x)\sin(m\varphi),\end{aligned} \tag{7}$$

where $J_m(x)$ is the Bessel function of the first kind of integer order $m$ and argument $x$, the terms $\cos\left[2k_p \sin(\omega_p l)\right]$ or $\sin\left[2k_p \sin(\omega_p l)\right]$ can be expressed as a series of harmonic sinusoidal components providing a cluster of symmetrically placed peaks in the spectrum with separation of angular frequency $\omega_p$ between adjacent peaks.

## 3. Model Parameter Estimation

The parameters of the modeled signal are estimated in the following sequence: First, we estimate the carrier and the modulating angular frequencies. At the next stage, the amplitude and the phase of the carrier, together with the modulation index of the PM signal are estimated. Finally, the phases, the scaling factors and the modulation index of the AM signal are estimated.

### 3.1 Estimation of Carrier and Modulating Angular Frequencies

The product function $p_x[l]$ is fitted into an autoregressive (AR) process, and the AR coefficients are calculated by using the modified covariance method, and in turn, the power spectral density (PSD) of $p_x[l]$ is computed by using the AR coefficients [12]. Alternatively, the AR coefficients can be used as the linear prediction (LP) coefficients to form the prediction error filter (PEF) whose roots are extracted to estimate the angular-frequency contents of the sequence $p_x[l]$. The order of an appropriate AR model is chosen to be sufficiently high so that the PSD computation does not miss any frequency of the product function $p_x[l]$.

As explained in the previous section, the PSD plot of $p_x[l]$ will have peaks at angular frequencies $\omega_a$ and $2\omega_a$, together with a dc component, and clusters of peaks centered at angular frequencies $2\omega_c$, $(2\omega_c - \omega_a)$, $(2\omega_c + \omega_a)$, $(2\omega_c - 2\omega_a)$ and $(2\omega_c + 2\omega_a)$. The highest and the second highest center frequencies of these clusters will be at $(2\omega_c + 2\omega_a)$ and $(2\omega_c + \omega_a)$ respectively. In the cluster of peaks, the center frequency can be identified using the following property of Bessel function [18]

$$J_n(x) = (-1)^n J_{-n}(x) \qquad (8)$$

which provides the symmetry in the heights of the side-peaks around the center frequency in the PSD plot.

Once $(2\omega_c + 2\omega_a)$ and $(2\omega_c + \omega_a)$ are known, then the carrier angular frequency $\omega_c$ and the modulating angular frequency $\omega_a$ for the AM signal can be computed. The separation of two adjacent peaks in any cluster is given by the modulating angular frequency $\omega_p$ for the PM signal. By taking an average of several such separations, the angular frequency $\omega_p$ can be accurately determined.

### 3.2 Estimation of Amplitude, Phase of Carrier and Modulation Index of PM Signal

Once the carrier and modulating angular frequencies are known, the amplitude and phase of the carrier, as well as, the modulating index of the PM signal can be estimated as explained in the sequel.

Given a set of discrete observation points $\{x[n]; \ n = 0, 1, \cdots, N-1\}$ of the signal to be modeled, we can rewrite (1) as

$$x[n] = \frac{A}{2} e^{j\theta} \sum_{m=-\infty}^{\infty} J_m(k_p) e^{j[\omega_c n + m\omega_p n]} + \frac{A}{2} e^{-j\theta} \sum_{m=-\infty}^{\infty} J_m(-k_p) e^{-j[\omega_c n - m\omega_p n]}$$

$$+ \frac{sAk_a}{4} e^{j[\theta+\theta_a]} \sum_{m=-\infty}^{\infty} J_m(k_p) e^{j[\omega_c n + \omega_a n + m\omega_p n]}$$

$$+ \frac{sAk_a}{4} e^{-j[\theta+\theta_a]} \sum_{m=-\infty}^{\infty} J_m(-k_p) e^{-j[\omega_c n + \omega_a n - m\omega_p n]} \qquad (9)$$

$$+ \frac{rAk_a}{4} e^{j[\theta-\theta_a-\theta_b]} \sum_{m=-\infty}^{\infty} J_m(k_p) e^{j[\omega_c n - \omega_a n + m\omega_p n]}$$

$$+ \frac{rAk_a}{4} e^{-j[\theta-\theta_a-\theta_b]} \sum_{m=-\infty}^{\infty} J_m(-k_p) e^{-j[\omega_c n - \omega_a n - m\omega_p n]}$$

where we use the identity $\cos[\varphi] = \left[ e^{j\varphi} + e^{-j\varphi} \right]/2$ together with (6) to expand the terms. Note that although each term of (9) contains theoretically an infinite number of side frequencies, in reality, there are only a finite number of significant side frequencies. The number of significant side frequencies varies with the modulation index $k_p$, and can be determined from tabulated values of the Bessel function $J_m(k_p)$ by setting an appropriate threshold [18].

Now, we define the complex amplitudes of (9) as follows

$$A_{c1} = \frac{A}{2} e^{j\theta}, \qquad A_{c2} = \frac{A}{2} e^{-j\theta},$$

$$A_{c3} = \frac{sAk_a}{4} e^{j[\theta+\theta_a]}, \qquad A_{c4} = \frac{sAk_a}{4} e^{-j[\theta+\theta_a]}, \qquad (10)$$

$$A_{c5} = \frac{rAk_a}{4} e^{j[\theta-\theta_a-\theta_b]}, \qquad A_{c6} = \frac{rAk_a}{4} e^{-j[\theta-\theta_a-\theta_b]}.$$

We also define the row vectors **W** and the column vectors **B** as shown below:

$$\mathbf{W}_{n,1} = \left[ e^{j[\omega_c n - M\omega_p n]}, e^{j[\omega_c n - (M-1)\omega_p n]}, \cdots, e^{j[\omega_c n + M\omega_p n]} \right],$$

$$\mathbf{W}_{n,2} = \left[ e^{-j[\omega_c n + M\omega_p n]}, e^{-j[\omega_c n + (M-1)\omega_p n]}, \cdots, e^{-j[\omega_c n - M\omega_p n]} \right],$$

$$\mathbf{W}_{n,3} = \left[ e^{j[\omega_c n + \omega_a n - M\omega_p n]}, e^{j[\omega_c n + \omega_a n - (M-1)\omega_p n]}, \cdots, e^{j[\omega_c n + \omega_a n + M\omega_p n]} \right], \quad (11)$$

$$\mathbf{W}_{n,4} = \left[ e^{-j[\omega_c n + \omega_a n + M\omega_p n]}, e^{-j[\omega_c n + \omega_a n + (M-1)\omega_p n]}, \cdots, e^{-j[\omega_c n + \omega_a n - M\omega_p n]} \right],$$

$$\mathbf{W}_{n,5} = \left[ e^{j[\omega_c n - \omega_a n - M\omega_p n]}, e^{j[\omega_c n - \omega_a n - (M-1)\omega_p n]}, \cdots, e^{j[\omega_c n - \omega_a n + M\omega_p n]} \right],$$

$$\mathbf{W}_{n,6} = \left[ e^{-j[\omega_c n - \omega_a n + M\omega_p n]}, e^{-j[\omega_c n - \omega_a n + (M-1)\omega_p n]}, \cdots, e^{-j[\omega_c n - \omega_a n - M\omega_p n]} \right],$$

and,

$$\mathbf{B}_1 = \left[ J_{-M}(k_p), J_{-(M-1)}(k_p), \cdots, J_{(M-1)}(k_p), J_M(k_p) \right]^T,$$
$$\mathbf{B}_2 = \left[ J_{-M}(-k_p), J_{-(M-1)}(-k_p), \cdots, J_{(M-1)}(-k_p), J_M(-k_p) \right]^T, \quad (12)$$

where $M$ is the largest value of the integer $m$ that satisfies the requirement $\left| J_m(k_p) \right| > 0.01$ [18].

Then, writing the column vectors $\mathbf{R}$ to include the complex amplitudes,

$$\begin{aligned}
\mathbf{R}_1 &= A_{c1}\,\mathbf{B}_1, & \mathbf{R}_2 &= A_{c2}\,\mathbf{B}_2, \\
\mathbf{R}_3 &= A_{c3}\,\mathbf{B}_1, & \mathbf{R}_4 &= A_{c4}\,\mathbf{B}_2, \\
\mathbf{R}_5 &= A_{c5}\,\mathbf{B}_1, & \mathbf{R}_6 &= A_{c6}\,\mathbf{B}_2,
\end{aligned} \quad (13)$$

we can rewrite (9) in the matrix form as follows

$$\begin{bmatrix} x[0] \\ x[1] \\ \vdots \\ x[N-1] \end{bmatrix} = \begin{bmatrix} \mathbf{W}_{0,1} & \mathbf{W}_{0,2} & \mathbf{W}_{0,3} & \mathbf{W}_{0,4} & \mathbf{W}_{0,5} & \mathbf{W}_{0,6} \\ \mathbf{W}_{1,1} & \mathbf{W}_{1,2} & \mathbf{W}_{1,3} & \mathbf{W}_{1,4} & \mathbf{W}_{1,5} & \mathbf{W}_{1,6} \\ \vdots & \vdots & \vdots & \vdots & \vdots & \vdots \\ \mathbf{W}_{N-1,1} & \mathbf{W}_{N-1,2} & \mathbf{W}_{N-1,3} & \mathbf{W}_{N-1,4} & \mathbf{W}_{N-1,5} & \mathbf{W}_{N-1,6} \end{bmatrix} \begin{bmatrix} \mathbf{R}_1 \\ \mathbf{R}_2 \\ \mathbf{R}_3 \\ \mathbf{R}_4 \\ \mathbf{R}_5 \\ \mathbf{R}_6 \end{bmatrix} \quad (14)$$

which can be written as

$$\mathbf{X} = \overline{\mathbf{W}}\,\overline{\mathbf{R}}$$

where $\overline{\mathbf{W}}$ is a rectangular matrix of size $N \times L$, $L = 6(2M+1)$, and the value of $M$ is chosen to be sufficient to include all the significant side frequencies of the PM signal [18].

The solution of (14) gives us the column vector $\overline{\mathbf{R}}$. We consider the part $\mathbf{R}_1$ of $\overline{\mathbf{R}}$, and in order to determine the value of modulation index $k_p$ for the PM signal, the ratios of Bessel functions

$$\frac{J_0(k_p)}{J_1(k_p)}, \frac{J_0(k_p)}{J_{-1}(k_p)}, \frac{J_0(k_p)}{J_2(k_p)}, \frac{J_0(k_p)}{J_{-2}(k_p)},$$

etc., are computed. Now utilizing each ratio of Bessel functions of different orders and the same argument, the modulation index $k_p$ can be computed by looking into an appropriate table [17]. Similar procedure is repeated for the column vector $\mathbf{R}_2$, and an average value of $k_p$ is calculated.

We now consider the column vector $\mathbf{R}_1$ in order to find the values for phase of carrier signal $\theta$. We know from the properties of Bessel functions [18] that

$$\sum_{m=-\infty}^{\infty} J_m^2(x) = 1 \qquad \text{for all } x \tag{15}$$

The value of $\theta$ is given by the argument of sum of squares of all elements in the column vector $\mathbf{R}_1$, that is,

$$\theta = \frac{1}{2} \arg\left(A_{c1}^2\right) \tag{16}$$

where arg stands for argument. This procedure is repeated for all elements of the column vector $\mathbf{R}_2$, and an average value for the phase of the carrier is estimated. Once the values of modulation index for the PM signal and the phase of the carrier are known, the amplitude of the carrier can be determined from each element of the column vector $\mathbf{R}_1$ or $\mathbf{R}_2$, and then, the computed values are averaged.

## 3.3 Estimation of Phases, Scaling Factors and Modulation Index of AM Signal

The estimation of the phases, scaling factors and modulation index of the AM signal can be done using (9). However, the estimation carried out in this manner is not accurate. It is clear from (9) that the first and second terms in the right hand side of the equation has much larger contribution as compared to the remaining terms. This dominance of the first two terms does not allow the accurate estimation of the remaining terms involving the parameters of the AM signal. Since all the parameters that constitute the first two terms of (9) are known, we subtract these two terms from the data sequence $x[n]$. Let the modified data set be called $x_0[n]$. The scaling factor $r$ has been set at unity without loss of generality.

The modified data set $x_0[n]$ is represented by

$$x_0[n] = \frac{sAk_a}{4} e^{j[\theta+\theta_a]} \sum_{m=-\infty}^{\infty} J_m(k_p) e^{j[\omega_c n + \omega_a n + m\omega_p n]}$$

$$+ \frac{sAk_a}{4} e^{-j[\theta+\theta_a]} \sum_{m=-\infty}^{\infty} J_m(-k_p) e^{-j[\omega_c n + \omega_a n - m\omega_p n]}$$

$$+ \frac{rAk_a}{4} e^{j[\theta-\theta_a-\theta_b]} \sum_{m=-\infty}^{\infty} J_m(k_p) e^{j[\omega_c n - \omega_a n + m\omega_p n]}$$

$$+ \frac{rAk_a}{4} e^{-j[\theta-\theta_a-\theta_b]} \sum_{m=-\infty}^{\infty} J_m(-k_p) e^{-j[\omega_c n - \omega_a n - m\omega_p n]}$$

(17)

Writing (17) in the matrix form using (10) − (13), we get

$$\begin{bmatrix} x_0[0] \\ x_0[1] \\ \vdots \\ x_0[N-1] \end{bmatrix} = \begin{bmatrix} \mathbf{W}_{0,3} & \mathbf{W}_{0,4} & \mathbf{W}_{0,5} & \mathbf{W}_{0,6} \\ \mathbf{W}_{1,3} & \mathbf{W}_{1,4} & \mathbf{W}_{1,5} & \mathbf{W}_{1,6} \\ \vdots & \vdots & \vdots & \vdots \\ \mathbf{W}_{N-1,3} & \mathbf{W}_{N-1,4} & \mathbf{W}_{N-1,5} & \mathbf{W}_{N-1,6} \end{bmatrix} \begin{bmatrix} \mathbf{R}_3 \\ \mathbf{R}_4 \\ \mathbf{R}_5 \\ \mathbf{R}_6 \end{bmatrix}$$

(18)

or

$$\mathbf{X}_0 = \overline{\mathbf{W}}_0 \overline{\mathbf{R}}_0$$

where $\overline{\mathbf{W}}_0$ is a rectangular matrix of size $N \times L_0$, $L_0 = 4(2M+1)$. The solution of (18) gives us the column vector $\overline{\mathbf{R}}_0$.

We can estimate the value of phase $\theta_a$ using the same procedure that is used for phase of carrier by taking column vectors $\mathbf{R}_3$ and $\mathbf{R}_4$.
This is given by the relationship

$$\theta + \theta_a = \frac{1}{2}\arg\left(A_{c3}^2\right) \tag{19}$$

From this equation, the value of $\theta_a$ can be found as the value of $\theta$ is already known. We can now estimate the value of phase $\theta_b$ using the above arguments for column vectors $\mathbf{R}_5$ and $\mathbf{R}_6$. This is given by the expression

$$\theta - \theta_a - \theta_b = \frac{1}{2}\arg\left(A_{c5}^2\right) \tag{20}$$

From this equation, the value of $\theta_b$ can be found as the value of $\theta$ and $\theta_a$ are already known.

Next we determine the value of scaling factor $s$. This is done using the property of Bessel function given in (15) and the relationship given below:

$$s = \left(\frac{\left|A_{c3}^2\right|}{\left|A_{c5}^2\right|}\right)^{1/2} \tag{21}$$

where $|\cdot|$ denotes modulus operation. Similarly the value of $s$ is found using the values of $A_{c4}$ and $A_{c6}$. Note that the value of $r$ is set to be $r=1$.

The modulation index of AM signal is the last parameter remaining to be estimated. We once again use the Bessel function property given by (15) for this estimation. Here we will use the vector $\mathbf{R}_3$ to obtain $A_{c3}^2$, and the relationship given below for estimation of $k_a$:

$$\frac{sAk_a}{4} = \left(\left|A_{c3}^2\right|\right)^{1/2} \tag{22}$$

where $|\cdot|$ denotes modulus operation. Now the value of $k_a$ can be determined as values of $A$ and $s$ are already known. The average value of $k_a$ is found out by repeating this process with the column vectors $\mathbf{R}_4$, $\mathbf{R}_5$ and $\mathbf{R}_6$.

### 3.4 Estimation of Time-Varying Model

The above procedure of parameter estimation is repeated for sufficiently large number of data-blocks. Then using the estimated values of one parameter $\xi_i$ at the time indices $n_i$ which are assumed to be the center points of the data-blocks, a time-varying expression is calculated by fitting those values into a polynomial $\beta(n_i)$ of appropriately chosen degree [19]. By using the time polynomials of the parameters along with (1), the signal can be regenerated over the entire time duration. The simulations done based on this method are presented in the following section.

## 4. Simulation Study

The performance of the model discussed in Section 2 is demonstrated by simulation study in this section. First, we fit the model on the synthesized APMS signal. The accuracy of the proposed method of representation is verified by estimating parameters of the signal corrupted with noise. The model is then fitted on the sampled data of EEG signal.

### 4.1 Synthesized Signal

An APMS signal is generated using the model proposed with (1) in Section 2. The parameters used to generate the signal are shown in Table 1. The total number of samples being considered is 251. First, the product function $p_x[m]$ as defined in (4) is computed. The product function is fitted into an AR model by the modified covariance method and the power spectral density (PSD) plot is obtained [12]. The carrier and modulating frequencies are obtained from the symmetry of the PSD plot.

Once the carrier and modulating angular frequencies are known, the matrix equation (14) is formed. This equation is solved to find the modulation index of PM signal, amplitude and phase of the carrier. It has been explained that the accuracy of estimation for other parameters is not good and the original data set therefore needs to be modified. Hence, we subtract the term $A\cos\left[\omega_c n + k_p \sin(\omega_p n) + \theta\right]$ from the synthesized data at every value of the time index n. Next, the matrix equation (18) is formed and solved to get the values of the rest of the parameters.

The effect of noise on the estimation procedure is studied by adding the zero mean additive noise, setting the signal-to-noise ratio (SNR) to 10 dB and 20 dB. The estimated values of parameters under no-noise condition and noisy conditions are compared with the actual values for accuracy in Table 1. The estimation procedure is found to be robust. The comparison of original synthesized signal and regenerated signal shows that the model fitting is accurate. The signal plots are not shown here.

**Table 1 Signal Parameters: Actual and Estimated**

| Parameter | Actual value | Estimated value No noise | Estimated value SNR=20 dB | Estimated value SNR=10 dB |
|---|---|---|---|---|
| $\omega_c$ | 1.256637061 | 1.256637084 | 1.256787442 | 1.255825467 |
| $\omega_a$ | 0.785398163 | 0.785398047 | 0.784026090 | 0.779400787 |
| $\omega_p$ | 0.062831853 | 0.062828671 | 0.062914628 | 0.065887093 |
| $A$ | 3.0 | 3.000000790 | 3.011541395 | 3.337074884 |
| $\theta$ | 0.261799387 | 0.261800104 | 0.268723811 | 0.290493121 |
| $\theta_a$ | 0.392699081 | 0.392704277 | 0.402573581 | 0.492307913 |
| $\theta_b$ | 0.314159265 | 0.314152764 | 0.308876081 | 0.239993258 |
| $k_a$ | 0.5 | 0.499999570 | 0.496310066 | 0.426672826 |
| $k_p$ | 0.4 | 0.4 | 0.41 | 0.45 |
| $s$ | 0.6 | 0.600000576 | 0.612875446 | 0.710162158 |

## 4.3 Natural EEG Signal

The suitability of the model for a natural EEG signal is demonstrated. The EEG signal of a subject, 25 year old male, is used for this purpose. The data is sampled at 500 Hz. The analog signal is band-pass filtered between 1 Hz and 200 Hz, and then after sampling digitally filtered for passband between 0.2 Hz and 200 Hz. The EEG signal is plotted in Fig. 1.

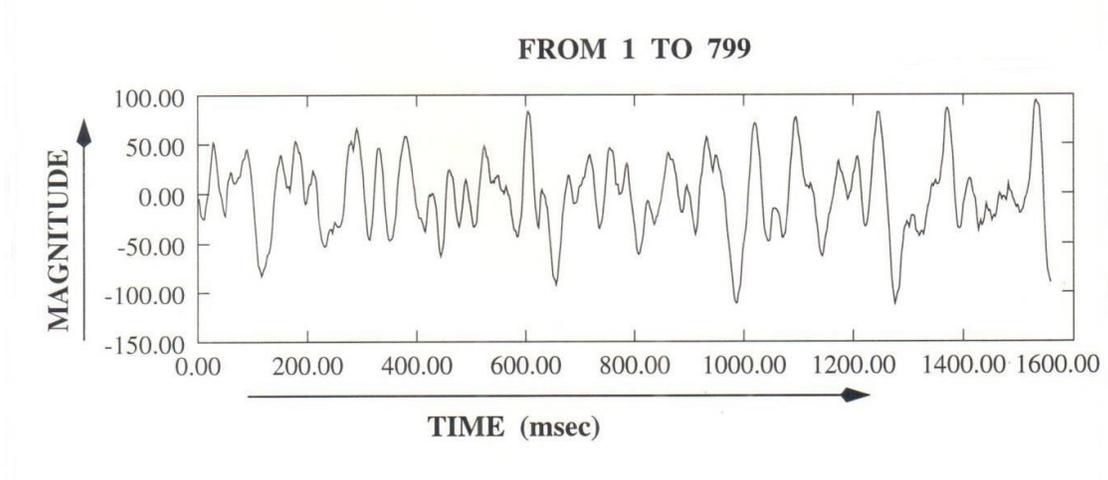

**Fig. 1 EEG Signal: 799 samples**

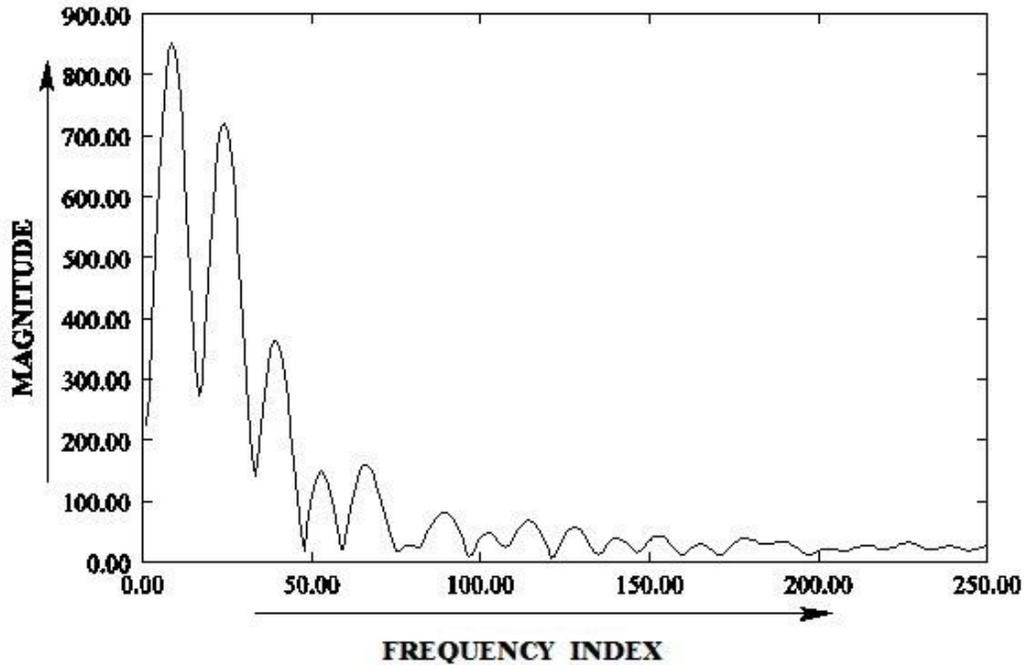

**Fig. 2 Discrete Fourier transform**

We carefully choose the number of sample points for a block of data for estimation of parameters of the model. The length of data block to be processed is decided by looking at the symmetry of three peaks at $(f_c - f_a)$, $f_c$ and $(f_c + f_a)$ of the discrete Fourier transform (DFT) plot as shown in Fig. 2. It is ensured that the modeled signal can be represented locally by the APMS signal when the symmetry of three peaks is obtained. Note that the modeled signal is a non-stationary signal with time-varying parameters. Therefore, the length of the data block is to be appropriately chosen such that the parameters do not vary appreciably over the block and the parameter estimation can be carried out with reasonable accuracy. In the simulation study, we consider each block consisting of 41 sample points. It turns out that the EEG signal to be modeled here is quite regular, and the DFT plot of each block has the desired symmetry.

The product function is computed using the chosen block of data. The parameters are estimated in the similar way as done for the synthesized signal. The original signal and corresponding regenerated signal for successive data-blocks are shown in Fig. 3 (i)–(xix).

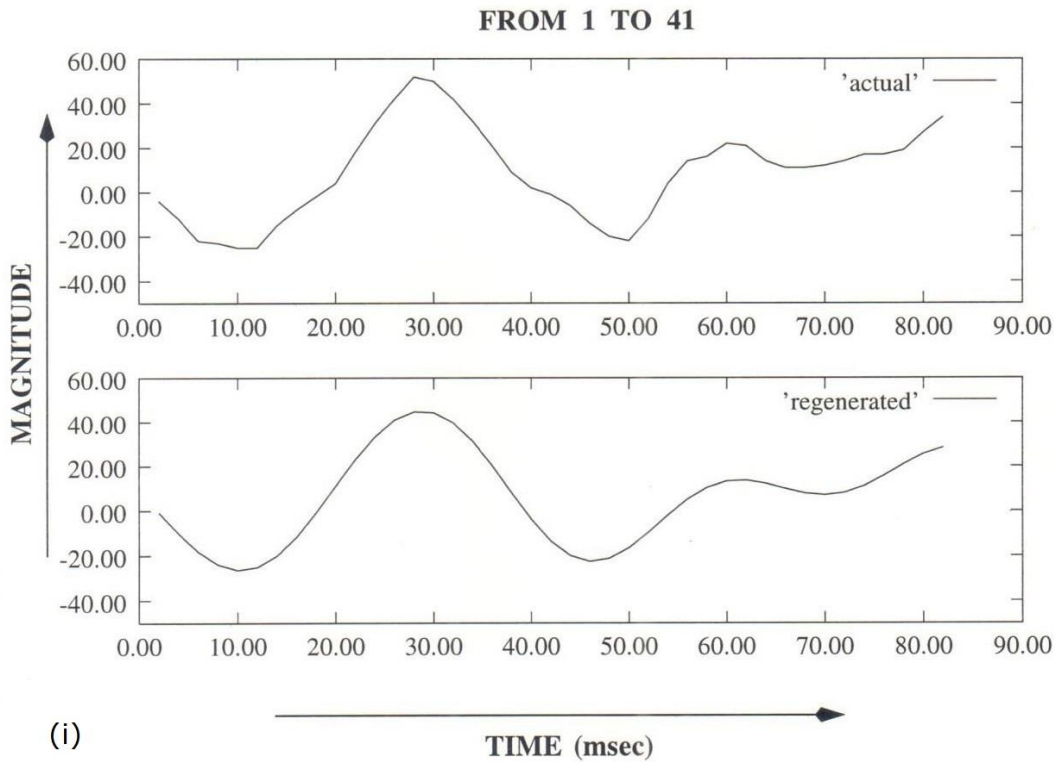

**Fig. 3 (i) Original and regenerated EEG signal segments**

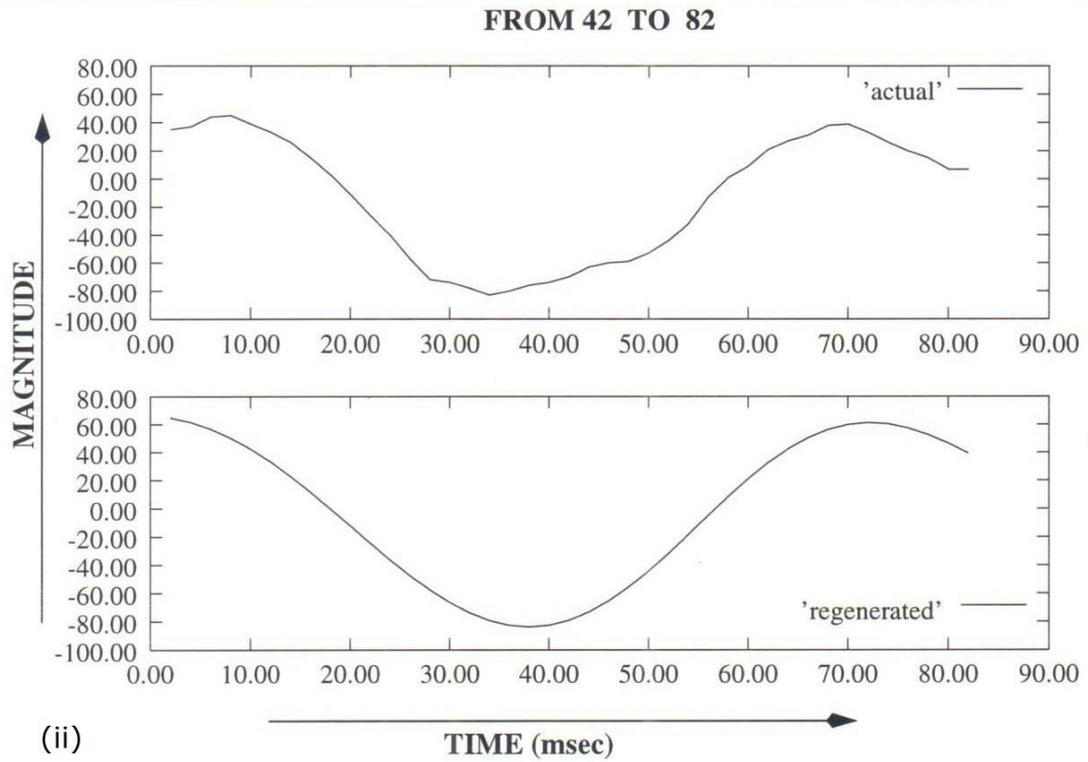

(ii)

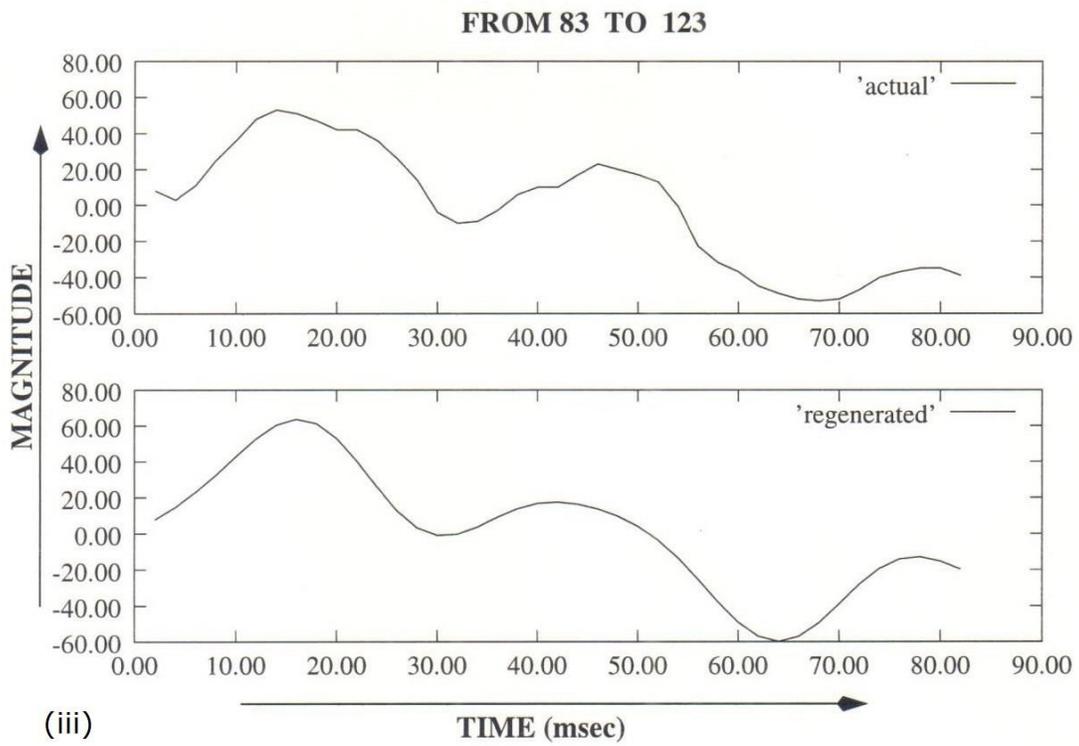

(iii)

**Fig. 3 (ii)(iii) Original and regenerated EEG signal segments**

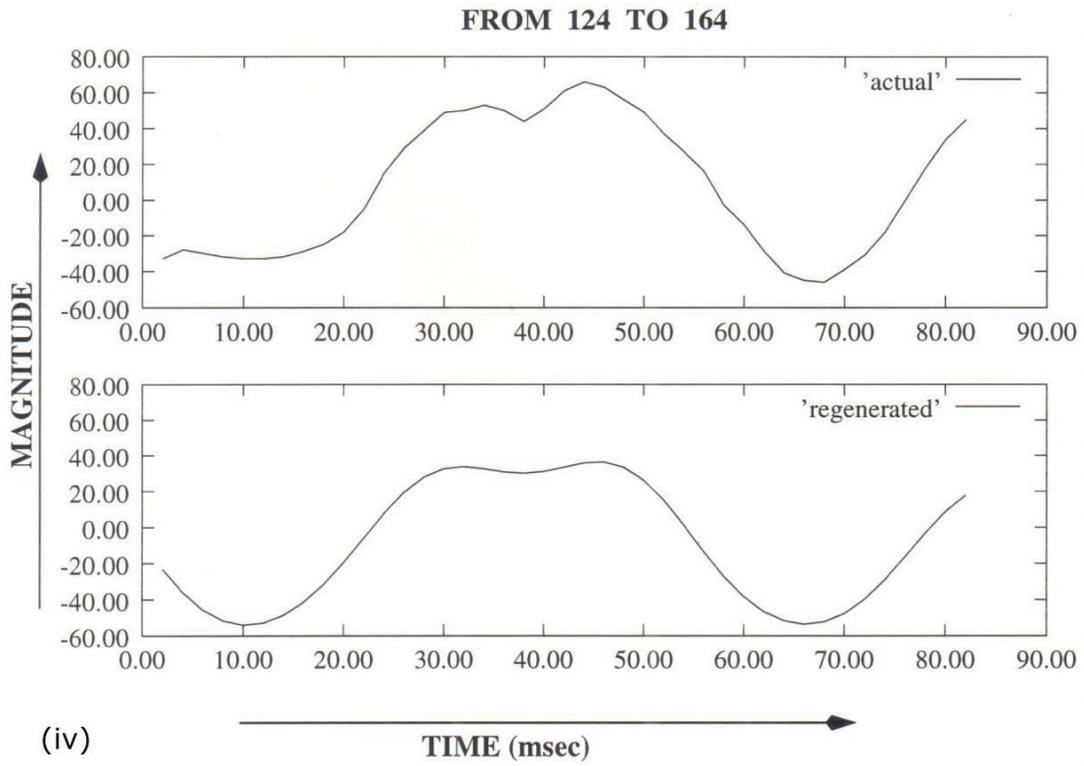

(iv)

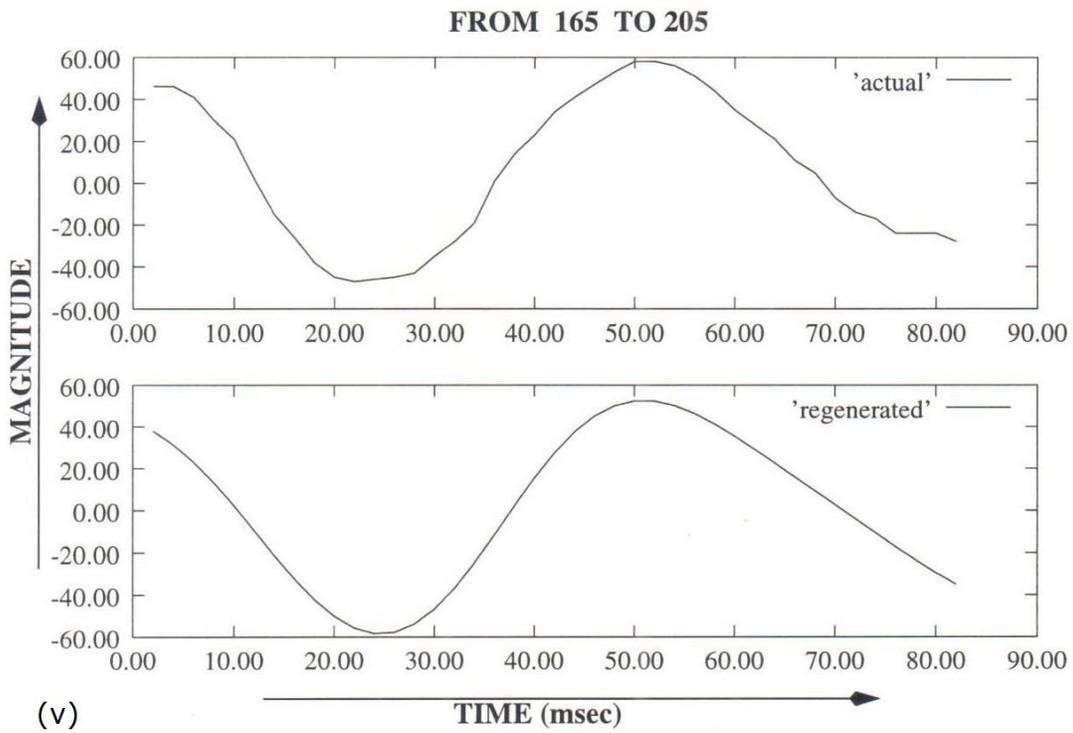

(v)

**Fig. 3 (iv)(v) Original and regenerated EEG signal segments**

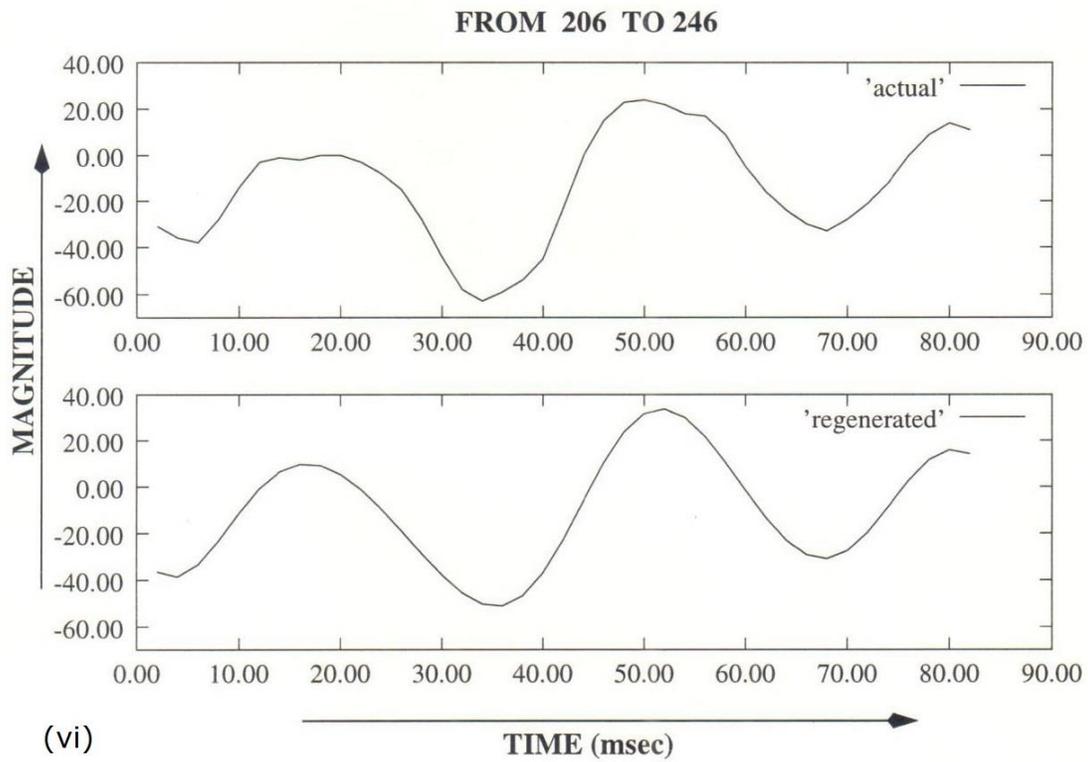

(vi)

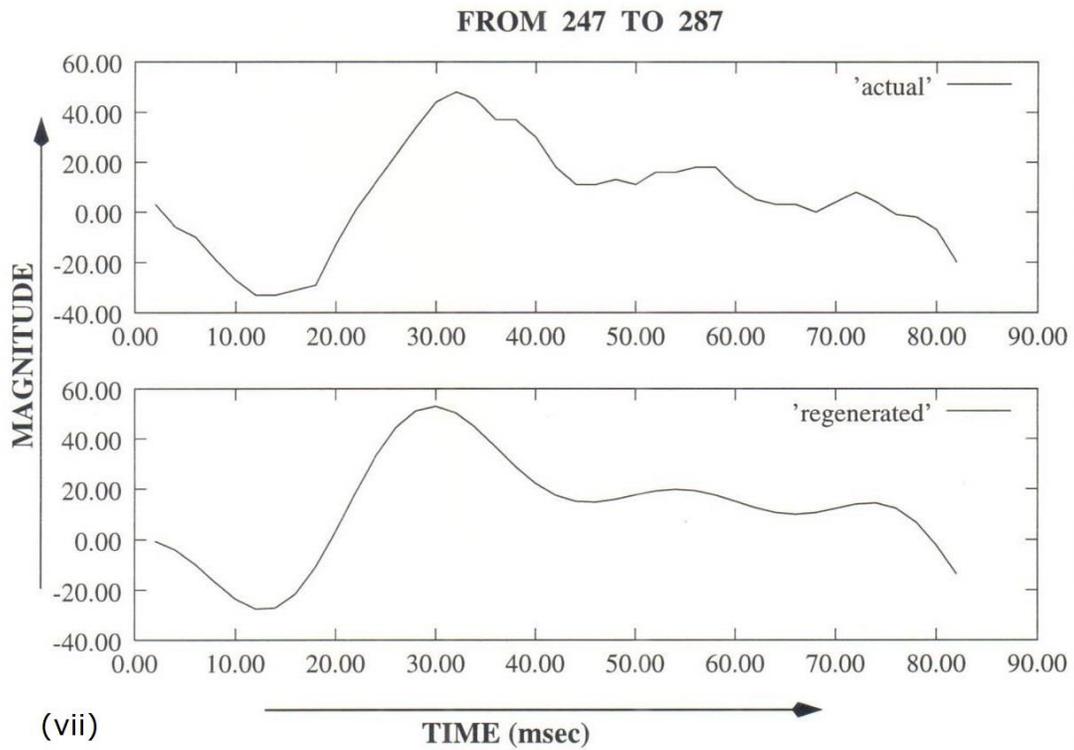

(vii)

**Fig. 3 (vi)(vii) Original and regenerated EEG signal segments**

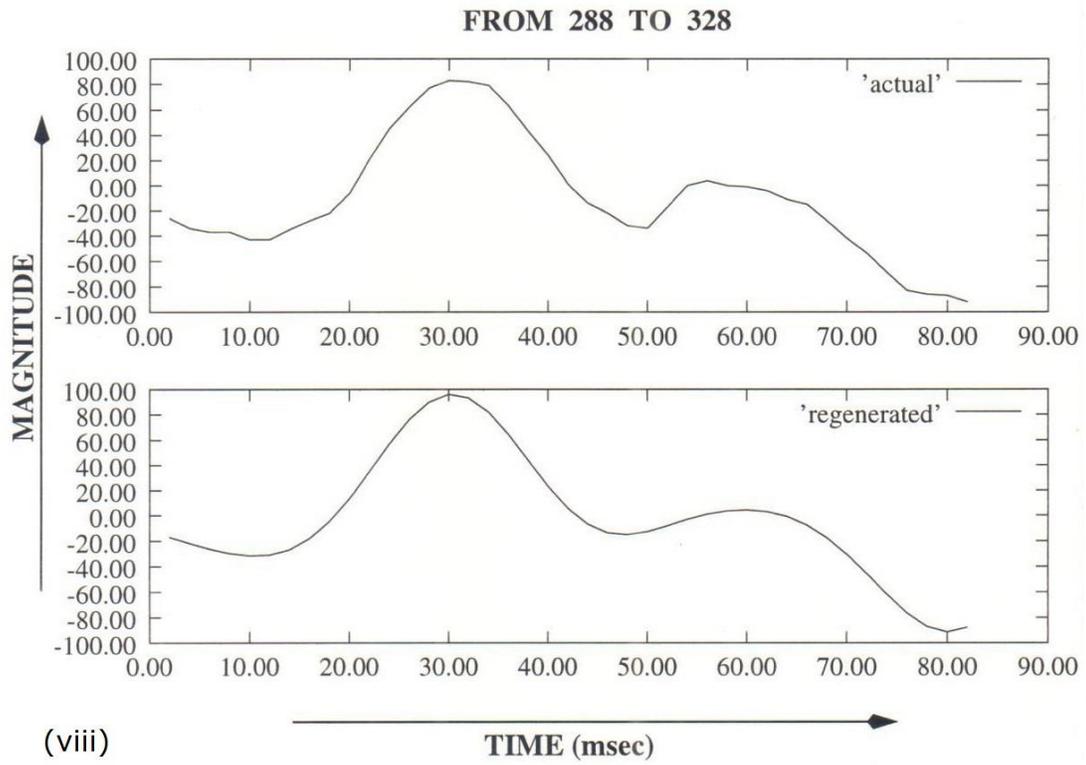

(viii)

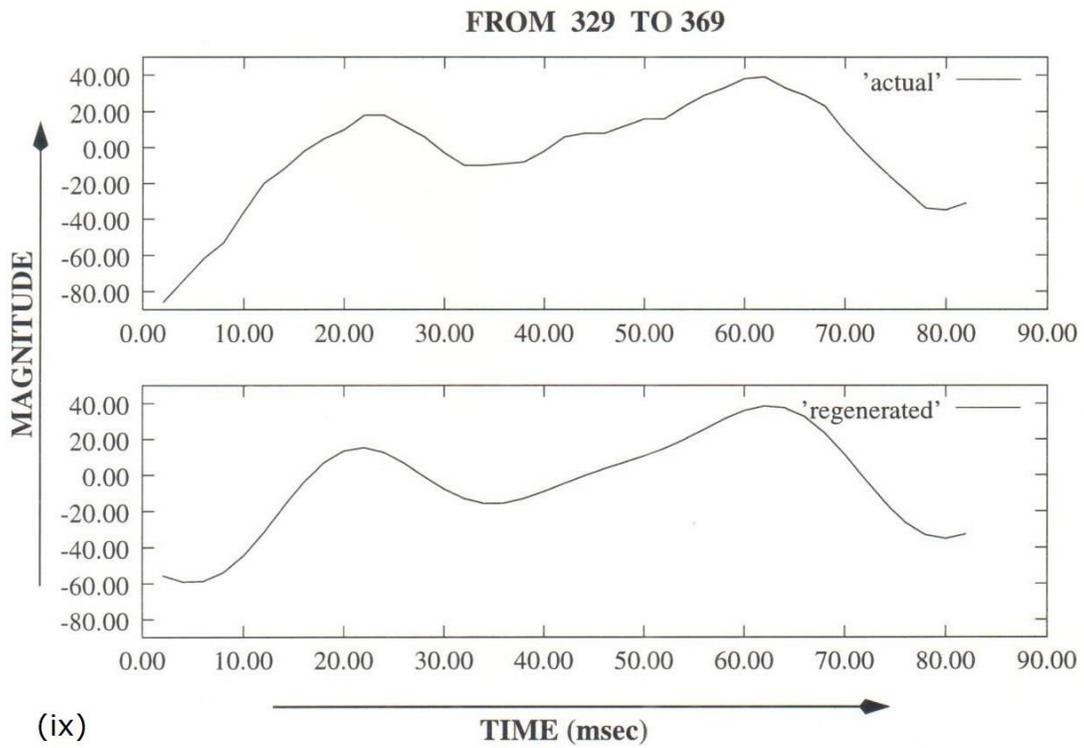

(ix)

**Fig. 3 (viii)(ix) Original and regenerated EEG signal segments**

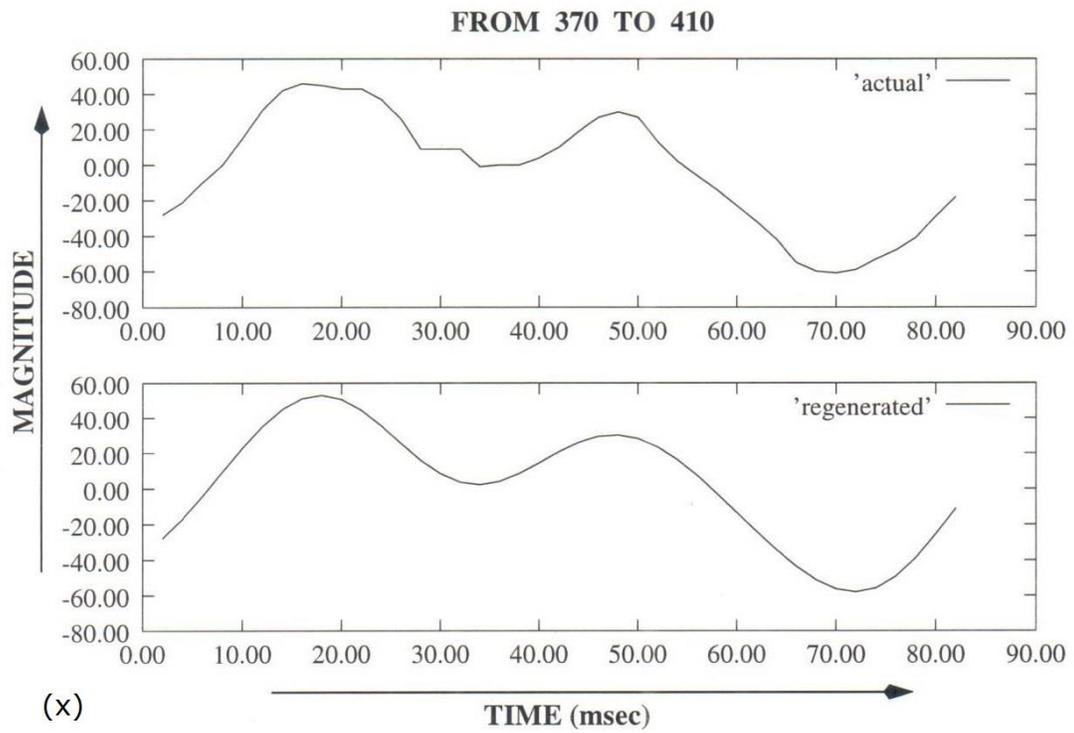

(x)

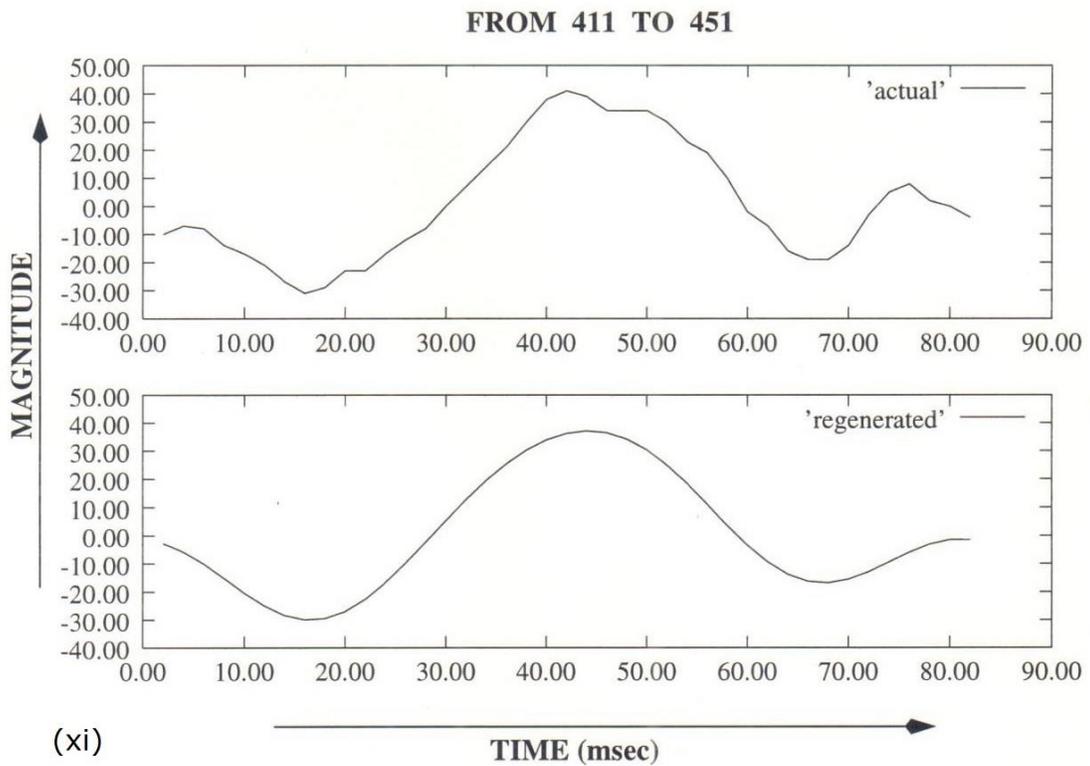

(xi)

**Fig. 3 (x)(xi) Original and regenerated EEG signal segments**

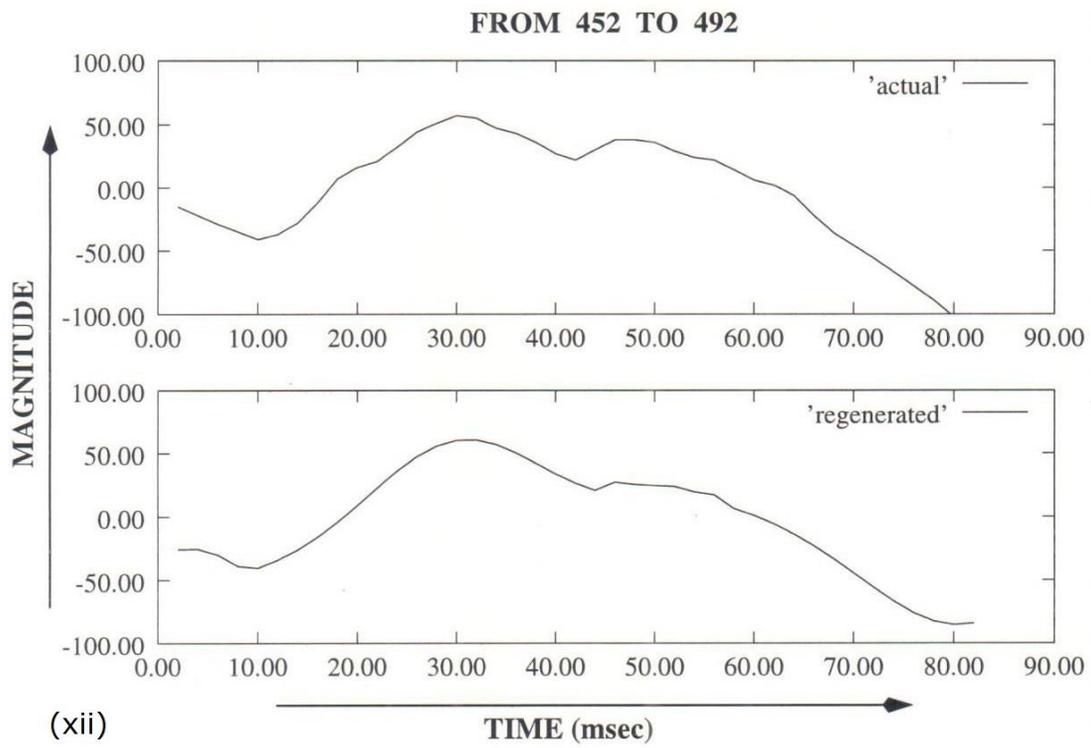
(xii)

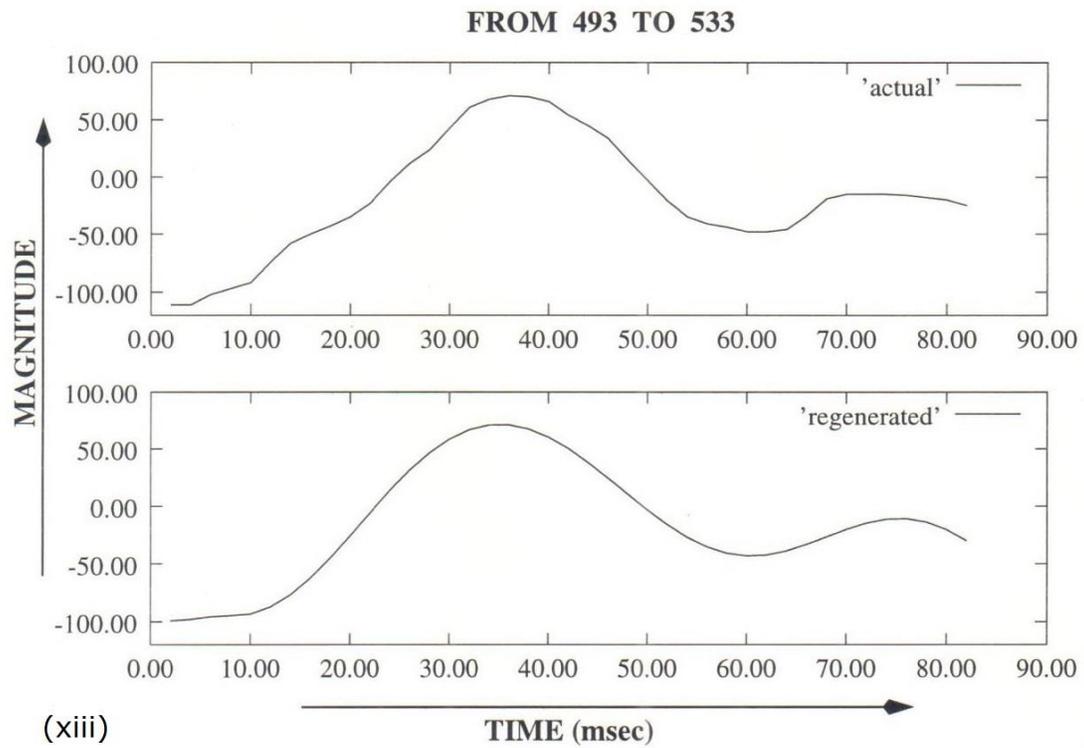
(xiii)

**Fig. 3 (xii)(xiii) Original and regenerated EEG signal segments**

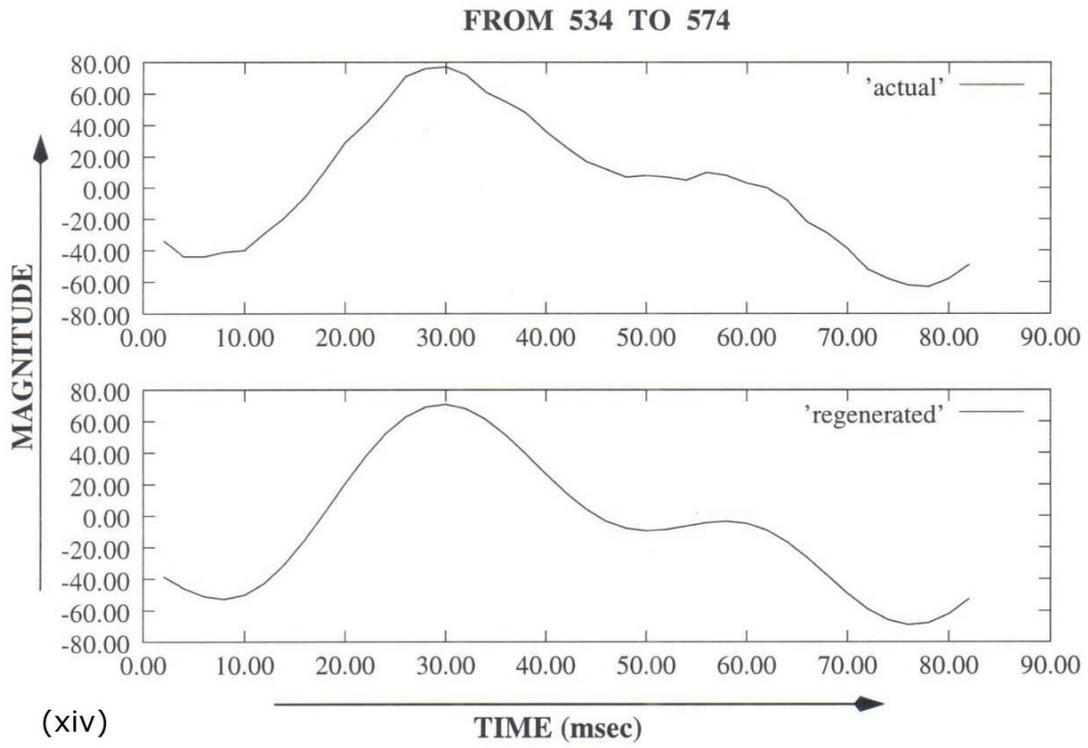

(xiv)

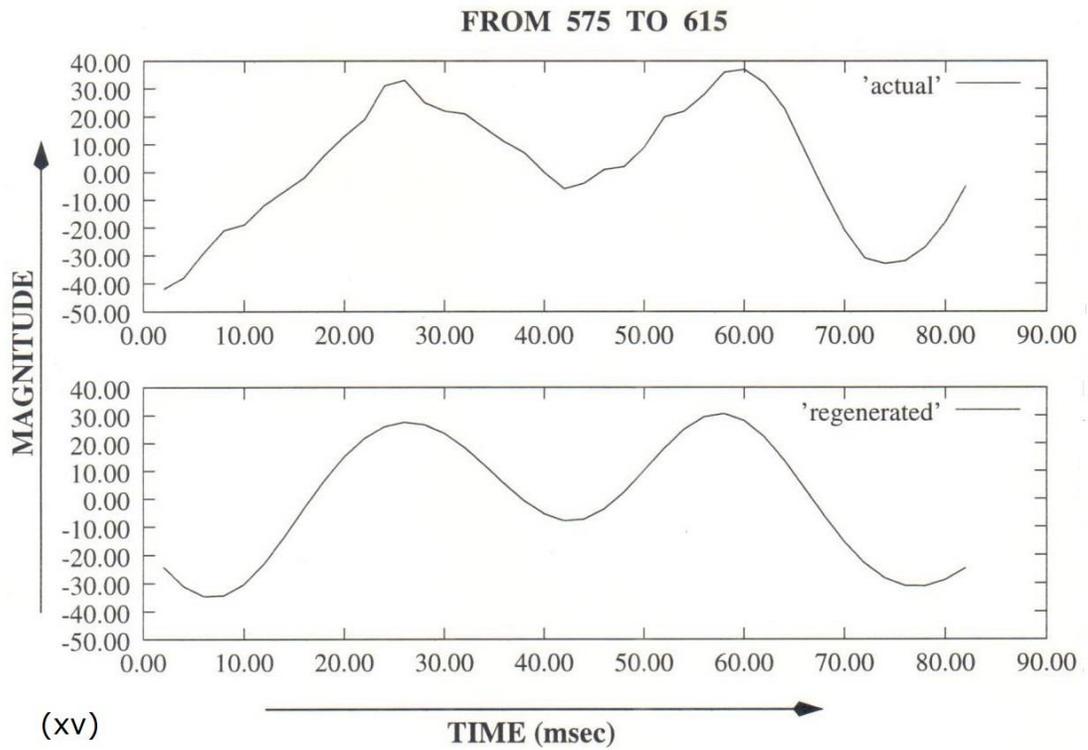

(xv)

**Fig. 3 (xiv)(xv) Original and regenerated EEG signal segments**

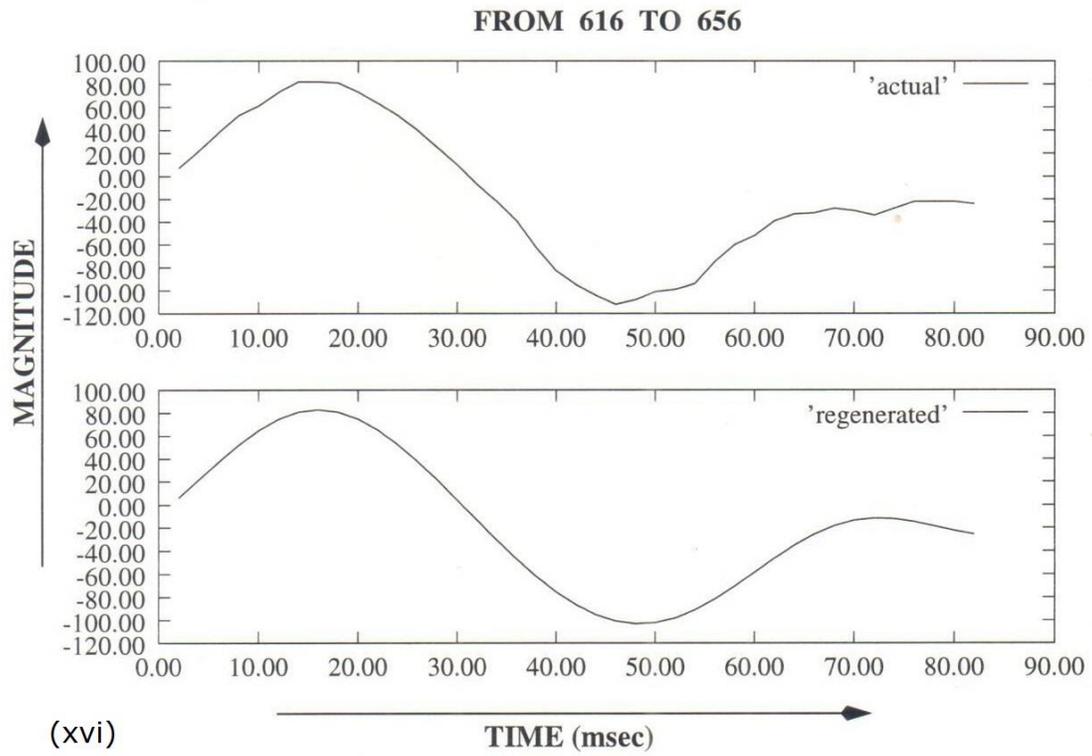

(xvi)

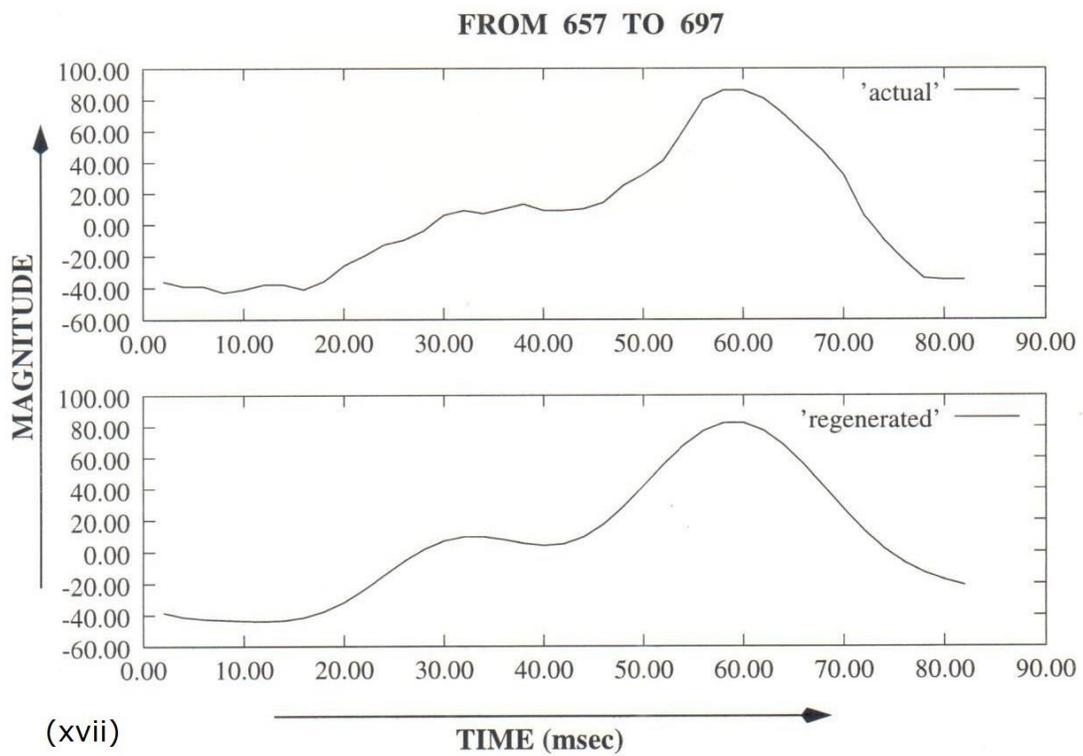

(xvii)

**Fig. 3 (xvi)(xvii) Original and regenerated EEG signal segments**

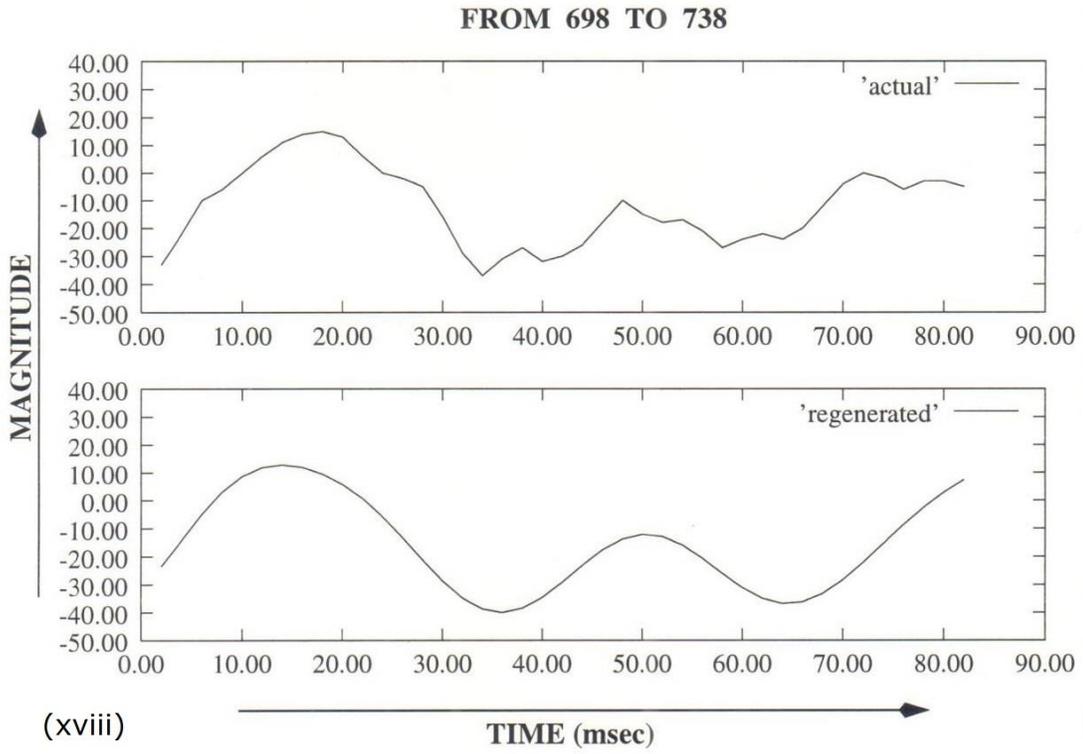

(xviii)

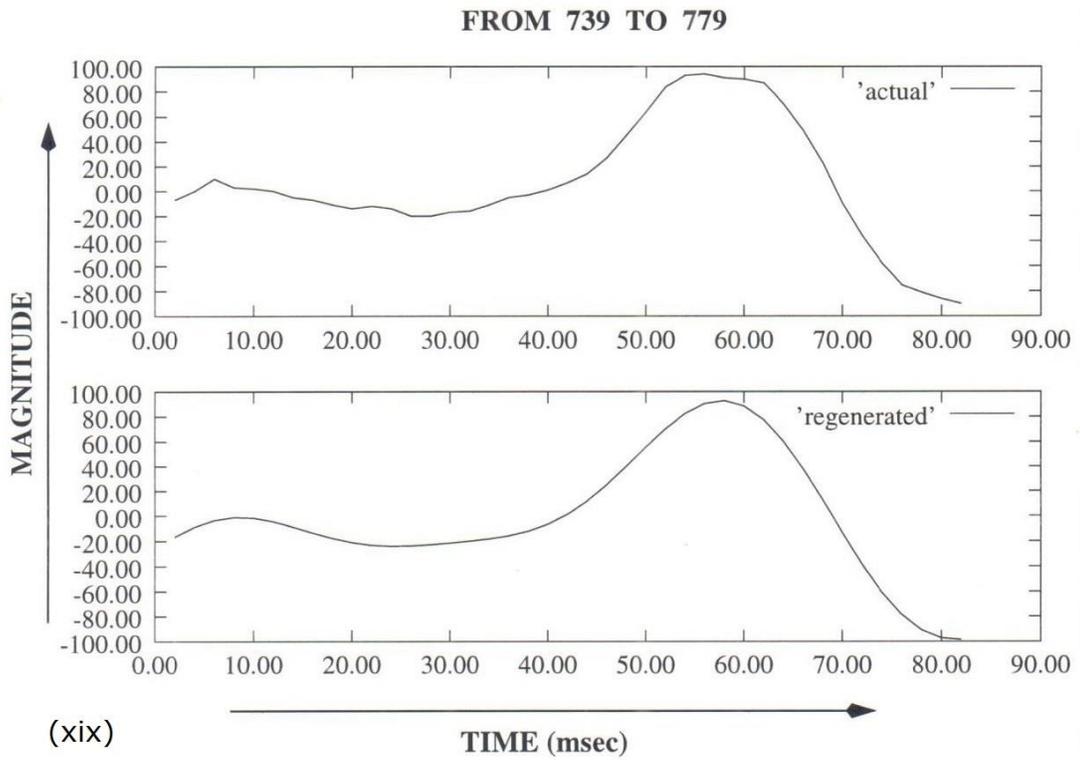

(xix)

**Fig. 3 (xviii)(xix) Original and regenerated EEG signal segments**

## 5. Conclusions

In this paper, we propose a new technique for parametric modeling of the EEG signal. The strategy developed for accurate estimation of parameters of the non-stationary model works in two stages: First, the signal is fitted locally into a simple model, and the parameters of the model are extracted. Next, the time variations of the parameters are estimated globally to represent the evolutionary nature of the signal.

We have demonstrated that the EEG signal with all its observed complex patterns can be represented by one single component amplitude and phase modulated signal, where the amplitude and phase are expressed as functions of time. We observe rhythmical activities of the EEG signal in some specific frequency ranges when the modulation laws are slowly varying with time, whereas transients, e.g., spikes or sharp waves, are observed in the EEG recording when the modulation laws vary rapidly with time.

The fact that the EEG signal is a single component non-stationary signal, as revealed in the present work, may have its own significance. Various non-stationary signal models developed so far, e.g., the complex exponential model for transient signals [20, 21], the complex AM-FM model for speech signals [17, 22, 23], and the exponential AM model for the ECG signal [24, 25] are all multi-component signal models. So it is amply surprising the EEG signal which appears to be quite complex in characteristic features, is indeed found to be a mono-component signal. It will be interesting to mention that our thought process, which is another projection of our brain activity, is a mono-event process as well, where one event can quickly change into another event, but there is no parallel processing.

Finally, we summarize our work by stating that the EEG signal is just one single signal which is evolutionary in nature. The signal manifests itself as various rhythms, spikes or sharp waves at different time spans. The realization that the diversified features of the EEG signal can be combined into a single-signal framework provides a higher level of understanding about the signal. How far we can make use of that concept to extract information depicted in the EEG signal will only be known by future research.